\documentclass{article}

\usepackage{arxiv}

\usepackage[utf8]{inputenc} 
\usepackage[T1]{fontenc}    
\usepackage{hyperref}       
\usepackage{url}            
\usepackage{booktabs}       
\usepackage{amsfonts}       
\usepackage{nicefrac}       
\usepackage{microtype}      
\usepackage{xspace}
\usepackage{array}
\usepackage{color, colortbl}
\usepackage{xcolor}
\usepackage{csquotes}
\usepackage[inline]{enumitem}
\usepackage{graphicx}

\definecolor{Gray}{gray}{0.9}
\newcommand{\servermix}{\textcolor{darkgray}{\textit{ServerMix}}\xspace}
\newcolumntype{L}[1]{>{\raggedright\let\newline\\\arraybackslash\hspace{0pt}}m{#1}}
 
\title{ServerMix: Tradeoffs and Challenges of Serverless Data Analytics
\thanks{
This work has been partially supported by the the European Union's Horizon 2020 research and innovation programme under grant agreement No 825184 and by the Spanish Ministry of Economy, Industry and Competitiveness through project ``Software Defined Edge Clouds'' (TIN2016-77836-C2-1-R).
}
}
\author{Pedro Garc\'{i}a-L\'{o}pez, Marc S\'{a}nchez-Artigas\\
Computer Engineering and Mathemathics Department\\
Universitat Rovira i Virgili\\
Tarragona, Spain \\
\texttt{(pedro.garcia,marc.sanchez)@urv.cat}\\
\And
Simon Shillaker, Peter Pietzuch\\
Large Scale Data and Systems Group\\
Imperial College London\\
London, England \\
\texttt{(s.shillaker17,prp)@imperial.ac.uk}\\
\And
David Breitgand, Gil Vernik\\
Cloud Platforms\\
IBM Research -- Haifa\\
Haifa, Israel \\
\texttt{(davidbr,gilv)@il.ibm.com}\\
\And
Pierre Sutra\\
CNRS, Universit\'e Paris Saclay, Télécom SudParis\\
\'Evry, France \\
\texttt{pierre.sutra@telecom-sudparis.eu}\\
\And
Tristan Tarrant\\
Red Hat\\
Cork, Ireland \\
\texttt{ttarrant@redhat.com}\\
\And
Ana Juan Ferrer\\
ATOS\\
Barcelona, Spain \\
\texttt{ana.juanf@atos.net}\\
}

\begin{document}
\maketitle

\begin{abstract}
Serverless computing  has become very popular today since it largely simplifies cloud programming. 
 Developers do not need to longer worry about provisioning or operating servers, and they pay only for 
the compute resources  used when their code is run.  This new cloud paradigm suits well for many applications, 
and researchers have already begun investigating the feasibility of serverless computing for data analytics. 
Unfortunately, today's serverless computing   presents important limitations that make it really difficult
to support all sorts of analytics workloads. This paper first starts by analyzing three fundamental trade-offs of
today's serverless computing model and their relationship with data analytics. It  studies how by relaxing disaggregation, isolation, 
and simple scheduling, it is possible to increase the overall computing performance,  but at the expense of essential
aspects of the model such as elasticity, security,  or sub-second activations, respectively. The consequence of these trade-offs is that
analytics applications may well end up embracing hybrid systems composed of serverless and serverful components, which we call \servermix in this paper.
We will review the existing related work to show that most applications can be actually categorized as \servermix.  Finally, this paper
will introduce the  major challenges of the CloudButton  research project to manage these trade-offs.
\end{abstract}

\keywords{Cloud computing \and Serverless computing \and Function-as-a-Service (FaaS)}

\section{Introduction}

With the emergence of serverless computing, the cloud has found a new paradigm that removes much of the complexity of  its usage by abstracting away the provisioning of compute
resources.  This fairly new model was culminated in $2015$ by Amazon in its Lambda service. This service  offered cloud  functions, marketed  as  FaaS  (Function  as  a  Service), 
and rapidly became the core  of  serverless computing. We say ``core'', because cloud  platforms  usually  provide  specialized  serverless  services  to  meet  specific application 
requirements, packaged as BaaS (Backend as a Service).  However, the focus of this paper will be on the FaaS model, and very often, the words ``serverless computing'' and 
``FaaS'' will be used interchangeably. The reason why FaaS drew widespread attention is because with FaaS platforms,  a user-defined function and its dependencies are deployed to the cloud,
where they are managed by the cloud provider and executed on demand. Simply put, users just write cloud functions in a high-level language and the serverless systems is who manages
everything else: instance selection, auto-scaling, deployment, sub-second billing, fault tolerance, and so on. The programming 
simplicity of functions  paves the way to soften the transition to the cloud ecosystem for end users.

Current practice shows that the FaaS model is well suited for many types of applications, provided that they require a small amount of storage and memory
(see, for instance, AWS Lambda operational limits~\cite{lambdaLimits}). Indeed, this model was originally designed to execute event-driven, stateless
functions in response to user actions or changes in the storage tier (e.g., uploading a photo to Amazon S$3$), which encompasses   many common tasks in
cloud applications. What was unclear is whether or not this new computing model could also be useful to execute data analytics applications. This question was answered
partially in $2017$ with the   appearance of two relevant research articles: ExCamera~\cite{ttt} and the  ``Occupy the Cloud'' paper~\cite{PyWren2017}. 
We say ``partially'' , because the workloads that both works handled mostly consisted of ``map''-only jobs,  just exploiting embarrassingly massive parallelism.
In particular,  ExCamera proved to be $60\%$ faster and $6$x cheaper than using VM instances when encoding vides on the fly over thousands of Lambda
functions.  The``Occupy the Cloud'' paper  showcased simple MapReduce jobs executed over Lambda Functions in their PyWren prototype. In this case, 
PyWren was $17\%$ slower than PySpark running on \texttt{r3.xlarge} VM instances.  The authors claimed that the simplicity of configuration and inherent 
elasticity of Lambda functions outbalanced the performance penalty.  They, however, did not compare the costs between their Lambda experiments against
an equivalent execution with virtual machines (VMs).

While both research works showed the enormous potential of serverless data analytics, today's serverless computing offerings  importantly restrict the ability to work efficiently with data. 
In simpler terms,  serverless data anlytics are way more expensive and less performant than cluster computing  systems or even VMs running analytics engines such as Spark.
The two recent articles~\cite{hellerstein2018serverless, berkeleyserverless} have outlined the major limitations of the serverless model in  general. Remarkably, ~\cite{berkeleyserverless} reviews the performance 
and cost of several data analytics applications, and shows that: a MapReduce-like  sort  of $100$TB  was $1\%$ faster than using VMs, but costing $15\%$ higher;
linear algebra computations~\cite{numpywren} were $3$x slower than an MPI implementation in a dedicated cluster, but only valid for large problem sizes;  
and machine learning (ML) pipelines were $3$x-$5$x faster than VM instances, but up to $7$x higher total cost.

Furthermore, existing approaches must rely on auxiliary serverful services to circumvent the limitations of the stateless serverless model.  For instance,
PyWren~\cite{PyWren2017} uses Amazon S$3$ for storage, coordination and as indirect communication channel. Locus~\cite{locus} uses Redis~\cite{redis}  through the ElastiCache service,
while  ExCamera~\cite{ttt} relies on a external VM-based rendezvous and communication service.  Also,  Cirrus~\cite{acaseML}  relies on disaggregated in-memory servers.

\subsection{On the path to serverless  data analytics:  the \servermix model}

In the absence of a fully-fledged serverless model in today's cloud platforms  (e.g., there is no effective solution to the question of serverles storage in the market),
current encarnations of serverless data analytics systems are  hybrid applications combining serverless and serverful services.   In this article, we identify 
them as ``\servermix''.  Actually, we will show how most related work can be classified under the  umbrella term of \servermix. 
 We will first  describe the existing design trade-offs involved in creating \servermix data analytics systems. 
We will then show that it is possible to relax core principles such as disaggregation, isolation, and simple scheduling to increase performance,
but also how this relaxation of the model may compromise the auto-scaling ability, security, and even the pricing model and fast startup time of serverless functions. 
For example:

\begin{itemize}
\item \textbf{Relaxation of disaggregation:}  Industry trends show a paradigm shift to disaggregated datacenters~\cite{Gao2016}. By physically  decoupling  resources and services,  datacenter  operators  can  easily customize their infrastructure to maximize the performance-per-dollar ratio.  One such example of this trend is serverless computing.  That is, FaaS offerings are of little value by themselves, and need of a vast ecosystem of dissagregated services
to build applications.  In the case of Amazon, this includes S$3$ (large object storage), DynamoDB (key-value storage), SQS (queuing services), SNS (notification services), etc. Consequently, departing from a serverless data-shipping model built around these services to a hybrid model where computations can be delegated to the stateful storage tier can easily achieve performance improvements~\cite{sampe2017data}. However, disaggregation is the fundamental pillar of  improved performance and elasticity in the cloud.
\item \textbf{Relaxation of  isolation:} serverless  platforms   leverage   operating   system containers  such as Docker to deploy and execute cloud functions.  In particular, each cloud function is hosted in a separate container.
However, functions of the same application may not need such a strong isolation and  be co-located  in the same container, which improves the performance of the application~\cite{sand}. Further, cloud functions are not directly 
network-addressable in any way.  Thus, providing direct communication between functions would reduce unnecessary latencies when multiple multiple functions interact with one another, such that one function's output is the input to another one.  Leveraging lightweight containers~\cite{oakes2018sock}, or even using language-level constructs would also reduce cold starts and boost inter-function communication.  However, strong isolation and sandboxing is the basis for multi-tenancy, fault isolation and security.

\item \textbf{Flexible QoS and scheduling:} current FaaS platforms only allow users to provision  some amount of RAM and a time slice of  CPU resources.  In the case of Amazon Lambda, the first determines the other. 
Actually, there is no way  to access specialized hardware or other resources such as the number of CPUs, GPUs, etc. To ensure service level objectives (SLOs), users should be able to specify resources requirements. But,
this would lead to implement complex scheduling algorithms that were able to reserve such resources, and even execute cloud functions in specialized hardware such as GPUs  with different isolation levels. However,
this would make it harder for cloud providers to achieve high resource utilization, as  more constraints are put on function scheduling. Simple user-agnostic scheduling is the basis for short start-up times and high
resource utilization, and should not be unwisely traded for the sole promise of better performance.  
\end{itemize}

It is clear that these approaches would obtain significant performance improvements. But, depending on the changes, such systems would be much closer to a serverful model based on VMs and dedicated resources than to 
the essence of serverless computing. In fact, we claim in this paper that the so-called ``limitations''of the serverless model are indeed its defining traits. When applications should require less disaggregation (computation close to the data), relaxation of isolation (co-location, direct communication), or tunable scheduling (predictable performance, hardware acceleration) a suitable solution is to build a \servermix solution. At least for serverless data analytics, 
we project that in a near future the dependency on  serverful computing will increasingly ``vanish'', for instance, by the appearance of high-throughput, low-latency BaaS storage services, 
so that many \servermix systems will eventually become $100\%$ serverless. Beyond some technical challenges, we do not see any fundamental reason why pure serverless data analytics would not flourish
in the coming years. 

In the meantime, we will scrutinize the \servermix model to provide a simplified programming environment, \textit{as much closer as possible to serverless},  for data analytics.  
To this aim, under the context of the H$2020$ CloudButton project, we will work  on the following three  points:  
\begin{enumerate*}[label=(\roman*)]
\item  Smart scheduling as a mechanism for providing transparent provisioning to applications while optimizing the cost-performance tuple in the cloud; 
\item Fine-grained mutable state disaggregation built upon consistent state services; and  
\item Lightweight and polyglot serverful  isolation: novel lightweight serverful FaaS runtimes based on WebAssembly~\cite{Haas2017} as universal multi-language substrate. 
\end{enumerate*}

\section{Fundamental trade-offs of serverless architectures}
\label{sec:trade-offs}

In this section, we will discuss three fundamental trade-offs underpinning cloud functions architectures —packaged  as  FaaS  offerings.  
Understand these trade-offs are important, not just for serverless data analytics, but to open the minds of designers to a broader range of 
serverless applications. While prior works  such as \cite{hellerstein2018serverless, berkeleyserverless} have already hinted these trade-offs,
the contribution of this section is to explain in more detail that the incorrect navigation of these trade-offs  can compromise essential aspects of the 
FaaS model. 

The first question is to ask is which are the essential aspects of the serverless model. For this endeavor, we will borrow the Amazon's definition of 
serverless computing, which is an unequivocal reference definition of this new technology.  According to this definition ~\cite{serverlessdef},  the four characteristic features of a serverless system are: 
\begin{itemize}
\item \textit{No server management:} implies that users do not  need to provision or maintain any servers;
\item \textit{Flexible scaling:} entails that the application can be scaled automatically via units of consumption (throughput, memory) rather than units of individual servers;
\item \textit{Pay for value:} is to pay for the use of  consumption units rather than server units; and
\item \textit{Automated high availability:} ensures that the system must provide  built-in availability and fault tolerance.
\end{itemize}

As we argue in this section, these four defining properties can be put in jeopardy but relaxing the tensions among three important architectural aspects that support them. 
These implementation aspects, which are \textit{disaggregation}, \textit{isolation}, and simple \textit{scheduling},  and their associated trade-offs, have major implications
on the success of the FaaS model.   In this sense, while a designer can decide to alter one or more of these trade-offs \textemdash for example, to improve performance, 
an oversimplifying or no comprehension of them can lead to hurt the four defining properties of the serverless model.  Let us see how the trade-offs affect them.

\subsection{Disaggregation} 

Disaggregation is the idea of decoupling resources over high bandwidth networks, giving us independent resource pools.
 Disaggregation has many benefits, but importantly, it allows each component to (auto-)scale in an independent manner.  
In serverless platforms,  disaggregation is the standard rather than an option, where  applications are run using stateless functions
that share state through  disaggregated storage (e.g., such Amazon S$3$)~\cite{hellerstein2018serverless, berkeleyserverless, numpywren}. 
This concept is backed up by the fact that modern high speed networks allow for sub-millisecond latencies  between the compute and storage layers 
\textemdash even allowing memory disaggregation like in  InfiniSwap \cite{gu2017efficient}.

Despite the apparent small latencies,  several works propose to relax disaggregation to favor performance. The reason is that
storage hierarchies,  across various storage layers and network delays, make disaggregation a bad design choice for
many latency and bandwidth-sensitive applications such as machine learning~\cite{acaseML}.  Indeed,~\cite{hellerstein2018serverless} 
considers that one of the limitations of serverless computing is its \textit{data-shipping architecture}, where data and state is regularly shipped to functions. 
To overcome this limitation, the same paper proposes the so-called ``\textit{fluid code and data placement}'' concept, where the infrastructure should be able to physically 
co-locate code and data. In a similar fashion,  \cite{serverlessos} proposes the notion of ``\textit{fluid multi-resource disaggregation}'', which consists of allowing movement 
(i.e., fluidity) between physical resources to enhance proximity, and thus performance. Another example of weakening disaggregation is  \cite{berkeleyserverless}. In this
paper, authors suggest to co-locate  related functions in the same VM instances for fast  data sharing. 

Unfortunately, while data locality reduces data movements, it can hinder the elastic scale-out of compute and storage resources.
 In an effort to scale out wider and more elastically, processing mechanisms near the data (e.g.,  active storage \cite{Active98})  have not 
been put at the forefront of cloud computing, though recently numerous proposals and solutions have emerged (see~\cite{Gustavo18} for details). 
Further, recent works such as \cite{sampe2017data} show that active storage computations can introduce resource contention and interferences into the storage service. 
For example, computations from one user can harm the storage service to other users, thereby increasing the running cost of the application (pay for value). In any case,
shipping code to data will interfere with the flexible scaling of serverless architectures due to the lack of  fast and elastic datastore in the cloud~\cite{acaseML}. 

Furthermore, ensuring locality for serverless functions would mean, for example, placing related functions in the same server or VM instance, while enabling fast shared memory between them. 
This would obviously improve performance in applications that require fast access to shared data such as machine learning and PRAM  algorithms, OpenMP-like implementations of
parallel algorithms, etc.  However, as pinpointed in \cite{berkeleyserverless}, besides going against the spirit of serverless computing, this approach would  reduce  the  flexibility  of  cloud  
providers  to  place  functions,  and  consequently  reduce  the capacity to scale-out while increasing the complexity of function scheduling. Importantly, this approach would force developers to think about 
low-level issues such as server management or whether function locality might lead sub-optimal load balancing  among server resources.

\subsection{Isolation} 

Isolation is another fundamental pillar of multi-tenant clouds services. Particularly, perfect isolation enables a cloud operator to run many functions (and applications) even on a single host, 
with low idle memory cost, and high resource efficiency.  What cloud providers seek is to reduce the overhead of multi-tenant function isolation  and provide high-performance (small startup times), 
for they leverage a wide variety of isolation technologies such as containers, unikernels, library OSes, or VMs. For instance, Amazon has recently released Firecracker microVMs~\cite{firecracker} for AWS Lambda and
Google has adopted  gVisor \cite{gvisor}. Other examples of isolation technologies for functions are CloudFlare Workers with WebAssembly \cite{cloudflare} or optimized containers such as SOCK \cite{oakes2018sock}.
These isolation techniques reduces startup times to the millisecond range,  as compared to the second timescale of traditional VMs.  Whether these lightweight solutions achieve parity to traditional VMs in terms of security remains 
to be shown.

Beyond the list of sandboxing technologies for serverless computing, most of them  battled-tested in the industry (e.g., Amazon Firecracker VMs), several research works have proposed to relax isolation in order
to improve performance. For instance, \cite{serverlessos} proposes  the  abstraction  of  a  process in serverless computing, with the property that each process can be run across multiple servers. As a consequence of
this multi-server vision, the paper introduces a new form of isolation that ensures  multi-tenant isolation across multiple servers (where the functions of the same tenant are run).  This new concept of isolation is called 
``\textit{coordinated isolation}'' in the paper. Further, \cite{hellerstein2018serverless} proposes two ways of relaxing isolation. The first one is  based on the ``\textit{fluid code and data placement}'' approach,
which has been described in the previous section. The second way is by allowing direct communication and network addressing between functions. In particular, the paper claims that today's serverless model stymies distributed computing due to its lack of direct communication amongst functions, and advocates for long-running, addressable virtual agents instead.

Another technique to increase performance is to relax isolation and co-locate functions in the same VMs or containers \cite{berkeleyserverless, sand}. Or even to provide very lightweight language-level constructs to reuse containers as much as possible.  This can make sense for functions belonging to the same tenant~\cite{sand}, since it would heavily reduce cold starts and execution time for function compositions (or workflows).  Unfortunately,
it is possible that independent sets of sandboxed functions compete for the same server resources and and  interfere  with  each  other’s  performance.  Or simply, that it becomes impossible to find a single host that
 have the necessary resources  for  running a sandbox of multiple functions, affecting important defining properties of serverless computing such as flexible scaling, pay for value and no server management, among others.

Experiencing similar issues as above,  it could be  also possible to enable direct communication between functions~of~the same tenant. In this case, direct communication would permit a variety of distributed communication models, allowing, for example, the construction of replicated shared memory between functions.

To put it baldly, each of these forms of relaxing isolation might in the end  increase the attack surface, for instance, by opening physical co-residency attacks and network attacks not just to single functions but a collection of them.

\subsection{Simple Scheduling} 

Simple scheduling is another essential pillar of serverless computing. Indeed, cloud providers can ensure Quality of Service (QoS) and Service Level Agreements (SLAs) to different tenants by appropriately scheduling the reserved resources and bill them correspondingly.   The goal of cloud scheduling algorithms is to maximize the utilization of the cloud resources while matching the requirements of the different tenants.

In today's FaaS offerings, tenants only specify the cloud function's memory size, while the function execution time is severely limited \textemdash for instance, AWS limits the execution time of functions to $15$ minutes. This single
constraint simplifies  the scheduling of cloud functions and makes it easy to achieve high resource utilization through statistical multiplexing. For many developers, this lack of control on specifying resources, such as the 
number of CPUs, GPUs, or other types of hardware accelerators, is seen as an obstacle.  To overcome this limitation,  a clear candidate would be to work on more sophisticated scheduling algorithms that support more constraints
on functions scheduling, such as hardware accelarators, GPUs, or the data dependencies between the cloud functions, which can lead to suboptimal function placement. For instance, it is not hard to figure out that a suboptimal placement
of functions can result in an excess of communication  to exchange data (e.g., for broadcast, aggregation and shuffle patterns~\cite{berkeleyserverless}) or in suboptimal performance. Ideally, these constraints should be 
(semi-)automatically inferred by the platform itself, for instance, from  static code analysis, profiling, etc., to not break the property of ``no server management '', i.e., the core principle of serverless. But even in this case, 
more constraints on function scheduling would make it harder to guarantee flexible scaling. 

The literature also propose ideas to provide predictable performance in serverless environments.  For instance, \cite{serverlessos} proposes the concept of  ``fine-grained live  orchestration'', 
which involves complex schedulers to allocate resources to  serverless processes that run across multiple servers in the datacenter.   \cite{hellerstein2018serverless} advocates for heterogeneous 
hardware support for functions where developers could specify their requirements in DSLs, and the cloud providers would then calculate the most cost-effective hardware to meet user SLOs. 
This would guarantee the use of specialized hardware for functions. In  \cite{berkeleyserverless},  it is supported the claim  to harness hardware heterogeneity in serverless computing. 
In particular, it is proposed that serverless systems could embrace multiple instance types (with prices according to hardware specs), or that cloud providers may select the hardware automatically depending 
on the code (like GPU hardware for CUDA code and TPU hardware for TensorFlow code).
  
Overall,  the general observation is that putting more constraints on function scheduling for performance reasons could be disadvantageous in terms of  flexible scaling, 
elasticity and even hinder high resource utilization.  Moreover, it would complicate the  pay per use model, as  it would make it difficult to pay for the use of  consumption units, 
rather than server units, due to hardware heterogeneity.

\subsection{Summary}

\begin{figure}[t]
  \centering
  \includegraphics[width=0.45\textwidth]{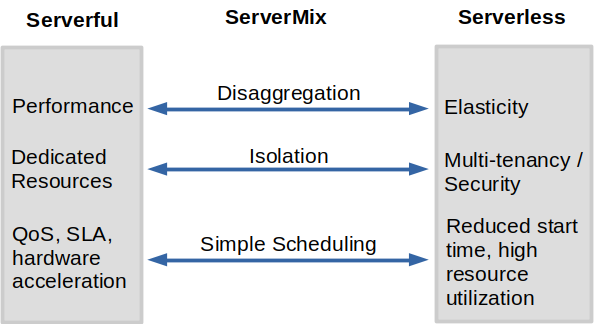}
  \caption{Tradeoffs}
  \label{fig:trade-offs}
\end{figure}

As a summary, we refer to Figure~\ref{fig:trade-offs} as a global view of the overall trade-offs.  These trade-offs have 
have serious implications on the serverless computing model and require careful examination. As we have already seen, 
disaggregation, isolation, and simplified scheduling are pivotal to ensure  flexible scaling, multi-tenancy, and millisecond 
startup times, respectively.

Weakening disaggregation to exploit function and data locality  can be useful to improve performance. However, it
 also means to  decrease the scale-out capacity of cloud functions and complicate function scheduling in order to meet
user SLOs.  The more you move to the left, the closer you are to serverful computing or running VMs or clusters in the datacenter.

With isolation the effect is similar. Since isolation is the key to multi-tenancy, completely relaxing isolation 
leaves nothing but dedicated resources. In your dedicated VMs, containers, or clusters (serverful), you can run functions very 
fast without caring about sandboxing and security. But this also entails more complex scheduling and  pricing models.

Finally, simple scheduling and agnostic function placement is also inherent to serverless computing. But if you require QoS, SLAs or specialized hardware, 
the scheduling and resource allocation gets more complex. Again, moved to the extreme, you end up in serverful settings that already exists (dedicated resources, VMs, or clusters).

Perhaps, the most interesting conclusion of this figure is the region in the middle, which we call \servermix computing. The zone in the middle involves applications that are built combining both 
serverless and serverful computing models. In fact, as we will review in the related work, many existing serverless applications may be considered \servermix according to our definition.

\section{Revisiting related work: The \servermix approach}
\label{sec:related-work}

\subsection{Serverless data analytics}  

Despite the stringent constraints of the FaaS model,  a number of works have managed to show how this model can be exploited to process and transform large amounts of 
data~\cite{PyWren2017,  ibmpywren, flint},  encode videos~\cite{ttt}, and run large-scale linear algebra computations~\cite{numpywren}, among other applications.  Surprisingly, and contrary to intuition, most 
of these  serverless data analytics  systems are indeed good \servermix examples, as they combine both serverless and serverful components.

In general, most of these systems rely on a external,  serverful provisioner component~\cite{PyWren2017,  ibmpywren, flint, ttt, numpywren}. This component is in charge of calling and orchestrating serverless functions using the APIs of the chosen cloud provider. Sometimes the provisioner is called ``coordinator''  (e.g.,  as in ExCamera~\cite{ttt}) or ``scheduler'' (e.g., as in Flint~\cite{flint}), but its role is the same: orchestrating  functions and providing some degree of fault tolerance. But the story does not end here.  Many of these systems require additional serverful components to overcome the limitations of the FaaS model. For example, recent works such as ~\cite{locus} use disaggregated in-memory systems such as ElastiCache Redis to overcome the throughput and speed  bottlenecks of slow disk-based storage services such as  S$3$. Or even external communication or coordination services to enable the communication among functions through a disaggregated intermediary (e.g., ExCamera~\cite{ttt}).

To fully understand the different variants of \servermix for data analytics, we will review each of the systems one by one in what follows. Table~\ref{tab1}  details which components
are serverful and serverless for each  system.
 
\begin{table}[htbp]
\def\arraystretch{1.5}
\caption{\servermix applications}
\begin{center}
\begin{tabular}{| p{2.8 cm} |L{4.5cm} |L{4.5cm} |}
\hline
\textbf{Systems}&\multicolumn{2}{c|}{\textbf{Components}} \\
\cline{2-3} 
&\textbf{\textit{Serverful}} & \textbf{\textit{Serverless} } \\\hline
\rowcolor{Gray}
PyWren~\cite{PyWren2017} & Scheduler  & AWS Lambda, Amazon S3   \\
IBM PyWren~\cite{ibmpywren} & Scheduler  & IBM Cloud Functions,  IBM COS, RabbitMQ   \\
\rowcolor{Gray}
ExCamera~\cite{ttt} & Coordinator and rendezvous servers (Amazon EC$2$ VMs) & AWS Lambda,  Amazon S3   \\
gg ~\cite{fouladi19}  & Coordinator & AW Lambda, Amazon S3,  Redis  \\
\rowcolor{Gray}
Flint~\cite{flint}  & Scheduler  (Spark context on client machine) & AW Lambda, Amazon S3, Amazon SQS   \\
Numpywren~\cite{numpywren} & Provisioner, scheduler (client process)  & AWS Lambda,  Amazon S3, Amazon SQS   \\
\rowcolor{Gray}
Cirrus~\cite{berkeleyserverless}  & Scheduler, parameter servers (large EC2   VM instances with GPUs) & AWS Lambda, Amazon S3  \\
Locus~\cite{locus}  & Scheduler, Redis service (AWS ElastiCache)  & AWS Lambda,  Amazon S3  \\
\hline
\end{tabular}
\label{tab1}
\end{center}
\end{table}

PyWren~\cite{PyWren2017} is a proof of concept that MapReduce tasks can be run as serverless functions. More precisely, PyWren consists of a serverful function scheduler (i.e., a client Python application) that permits to execute ``map'' computations  as AWS Lambda functions through a simple API.  The ``map'' code to be run in parallel is first serialized and then stored in Amazon S$3$. Next, PyWren invokes a common Lambda function that deserializes 
the ``map'' code and executes it on the  relevant  datum,   both  extracted  from S$3$.  Finally, the results are placed back  into S$3$. The scheduler actively polls S$3$  to detect that all partial results have been uploaded to S$3$ before
signaling the completion of the job.

IBM-PyWren~\cite{ibmpywren} is a PyWren derived project which adapts and extends PyWren for IBM Cloud services. It includes a number of new features, such as broader MapReduce support, automatic data discovery and partitioning, integration with Jupiter notebooks, and simple function composition, among others. For function coordination,  IBM-PyWren uses RabbitMQ to avoid the unnecessary  polling to the object storage service
(IBM COS), thereby improving job execution times compared with PyWren. 

ExCamera~\cite{ttt} performs digital video encoding by leveraging the parallelism of thousands of Lambda functions. Again, ExCamera uses serverless components (AWS Lambda, Amazon S3) and serverful ones (coordinator and  rendezvous servers). In this case, apart from a coordinator/scheduler component that starts and coordinates functions, ExCamera also needs of a rendezvous service, placed in an EC$2$ VM instance, to communicate functions
amongst each other.

Stanford's gg~\cite{fouladi19}  is an orchestration framework for building and executing burst-parallel applications over Cloud Functions. gg presents an intermediate representation that abstracts the compute and storage platform, and it provides dependency management and straggler mitigation. Again, gg relies on an external coordinator component, and an external Queue for submitting jobs (gg's thunks) to the execution engine (functions, containers). 

Flint \cite{flint} implements a serverless version of the PySpark MapReduce framework.  In particular, Flint replaces Spark executors by Lambda functions.  It is similar to PyWren in two main aspects. On one hand, it uses an external  serverful scheduler for function orchestration. On the other hand, it leverages S3 for input and output data storage. In addition, Flint uses  the Amazon's SQS service to store intermediate  data and perform the necessary data shuffling  to implement many of the PySpark's transformations.

Numpywren~\cite{numpywren} is a serverless system for executing large-scale dense linear algebra programs. Once again, we observe the \servermix pattern in numpywren. As it is based in PyWren, it relies on a external scheduler and Amazon S$3$ for input and ouput data storage. However, it adds an extra serverful component in the system called provisioner. The role of the provisioner is to monitor the length of the task queue and  increase the number of
Lambda functions (executors)   to match the dynamic parallelism during a job execution. The task queue is implemented using Amazon SQS. 

Cirrus machine learning (ML) project \cite{berkeleyserverless} is another example of a hybrid system that combines serverful components (parameter servers, scheduler) with serverless ones (AWS Lambda, Amazon S3).
 As with linear algebra algorithms,  ML researchers have traditionally used clusters of VM instances for the different tasks in ML workflows. Nonetheless, a fixed a cluster size can either lead to severe underutilization or  
slowdown, since each stage  of  a workflow can  demand significantly  different  amounts  of resources. Cirrus addresses this challenge by enabling every stage to scale to meet its resource demands by using Lambda functions.
The  main problem with Cirrus is that many ML algorithms require state to be shared between cloud functions, for it uses VM instances to share and store intermediate state. This
necessarily converts Cirrus into another example of a \servermix system.

Finally, the most recent example of \servermix is Locus~\cite{locus}. Locus targets one of the main limitations of the serverless stateless model: data shuffling and sorting. Due to the impossibility of function-to-function
communication, shuffling is ill-suited for serverless computing, leaving no other choice but to implement it through  an intermediary cloud service, which could be  cost prohibitive to deliver good performance. 
Indeed, the first attempt to provide an efficient shuffling operation was realized in PyWren~\cite{PyWren2017}  using $30$ Redis ElastiCache servers, which proved to be a very expensive solution. The major contribution of Locus was the development of  a hybrid  solution that considers both cost and performance.  To achieve an optimal cost-performance trade-off, Locus combined a small number of expensive fast Redis instances~with the much cheaper S3 service,
 achieving comparable performance to running Apache Spark on a provisioned cluster.

We did not include SAND \cite{sand} in the list of  \servermix systems. Rather, it  proposes a new FaaS runtime. In the article, the authors of SAND  present it as an alternative high-performance serverless platform. To deliver 
high performance,  SAND introduces two relaxations in the standard serverless model: one in \textit{disaggregation}, via a hierarchical message bus that enables function-to-function communication, 
and another in \textit{isolation}, through application-level sandboxing that enables packing multiple application-related functions together into the same container. Although SAND was shown
to deliver superior performance than Apache OpenWhisk, the paper failed to evaluate how these relaxations~can affect the scalability, elasticity and security of the standard FaaS model.

Recent works also outline the need for novel serverless  services providing flexible disaggregated storage to serverless functions. This is the case of Pocket's  ephemeral storage service~\cite{pocket}, which provides auto-scaling and pay-per-use as a service to cloud functions.  Similarly,  \cite{berkeleyserverless} proposes as a future challenge the creation of high-performance, affordable, transparently provisioned storage. This work discusses two types of 
unaddressed storage needs: \textit{Serverless Ephemeral Storage} and \textit{Serverless Durable Storage}, both of which should deliver micro-second latencies, fault-tolerance, auto-scalability and  transparent provisioning
for  multi-tenant workloads. The paper suggests that with a shared in-memory service,  spare memory resources  from one serverless application can be allocated to another. Finally, it also elaborates on why existing cloud services such as Redis or MemCached cannot fulfill the aforementioned storage needs. Actually, both in-memory services can be deemed as \textit{serverful} due to their need for explicit provisioning and dedicated resources per tenant.

\subsection{Serverless orchestration systems}  

 An interesting alternative could be to use serverless orchestration services to coordinate and provision data analytics applications. The key
question is whether these systems are actually well-prepared to orchestrate massively parallel computations.  In this sense, our recent paper~\cite{comparison} compares three commercial serverless orchestration services
along several axis: Amazon Step Functions, Azure Durable Functions, and IBM Composer. The  conclusion of our study was  nothing but pessimistic: all the alluded services showed a poor support~for parallelism.
However, a little time after the publication of our study, IBM Composer  improved their support for parallel function execution, so we believe it could become a solid alternative in a near future.

Our major observation is that if data analytics applications can some day be built atop of  serverless orchestration and  in-memory services with not much effort,  Table~\ref{tab1} could change importantly 
since many projects could avoid the use of serverful entities. This would clearly create more native applications with stateful and fault-tolerance models entirely provided in the cloud.  

Unfortunately, as of today, the reality is that cloud applications are still inadvertently bound to the \servermix model.  In fact, several cloud providers are now transitioning to hybrid models combining serverless and serverful concepts. For instance, we have recently seen how some cloud providers like Microsoft  offer \servermix FaaS services such as the Azure Premium Plan for running functions on dedicated machines while abstracting users from the provisioning phase.  The Azure platform is even allowing users to pre-warm functions  to reduce their cold starts.

\subsection{Serverless container  services} 

Hybrid cloud technologies are also accelerating the combination of serverless and serverful components. For instance, the recent deployment of Kubernetes (k$8$s) clusters in the big cloud 
vendors can~help overcome the existing application portability issues in the cloud.  There exists a plenty of hosted k$8$s services such as \textit{Amazon Elastic Container Service} (EKS),  \textit{Google Kubernetes Engine} (GKE), and \textit{Azure Kubernetes Service} (AKS), which confirm that this trend is gaining momentum.    However, none of these services can be considered $100\%$ ``serverless''.  Rather, they should be viewed as a middle ground between cluster computing and serverless computing. That is, while these hosted services offload operational management of  k$8$s,  they still require custom configuration by developers. The major similarity   to serverless computing is that 
k$8$s can provide short-lived computing environments like in the customary FaaS model.

But a very interesting recent trend is the emergence of the so-called serverless container services such as AWS Fargate, Azure Container Instances (ACI), and Google Cloud Run (GCR). These services reduce the complexity of managing and deploying k$8$s clusters in the cloud.  While they offer serverless features such as flexible automated scaling and pay-per-use billing model,  these services still  require some manual  configuration of the 
right parameters for the containers (e.g., compute, storage, and networking)  as well as the scaling limits for a successful deployment.

These alternatives are interesting for long-running jobs such as batch data analytics, while they offer more control over the applications thanks to the use of containers instead of functions. In any case, they can be very suitable for stateless, scalable applications, where the services can scale-out by easily adding or removing container instances.  In this case, the user establishes a simple CPU or memory threshold and the service is responsible for monitoring, load balancing, and instance creation and removal.  It must be noted that if the service or application is more complex (e.g., a stateful storage component), the utility of these approaches is rather small, or they require important user intervention.

For example, AWS Fargate offers two models: Fargate launch type and EC2 launch type. The former is more serverless and requires less configuration. The latter gives users more control but also more responsibility. 
An analogous issue occurs with Google: Cloud Run vs. Cloud Run on GKE. The former is automated and uses standard vCPUs, while the latter enables customers to select hardware requirements and manage their cluster.

An important open source project related to serverless containers is  CNCF's KNative. In short, KNative is backed by big vendors such as Google, IBM and RedHat, among others, and it simplifies the creation of serverless containers over k$8$s clusters. Knative simplifies the complexity of k$8$s and Istio service mesh components, and it creates a promising substrate for both PaaS and FaaS applications. GCR is based on Knative.  IBM Cloud is also offering seamless Knative integration in their k$8$s services.  Yet, it is hard to see how,  in future terms,  hosted KNative and k$8$s cloud services will reshape the current FaaS landscape, since, in its present form, have important implications
on the key traits of  the FaaS model such as disaggregation and scheduling. 
\bigskip

\textbf{As a final conclusion}, we foresee that the simplicity of the serverless model will gain traction among users, so many new offerings may emerge in the next few years, thereby blurring the borders between both serverless and serverful models. Further, container services may become an interesting architecture for \servermix deployments.

\section{CloudButton:  Towards serverless data analytics }

We strongly believe that serverless technologies can be a key enabler for radically-simpler, user-friendly data analytics systems in the coming years. However,
achieving  this goal requires a programmable framework that goes beyond the current FaaS model and has user-adjustable settings to alter the IDS (Isolation-Dissagregation-Scheduling) trade-off 
(see \S\ref{sec:trade-offs} for more details) \textemdash for example,  by weakening function isolation for better performance.

The EU CloudButton project~\cite{cloudbutton} was born out of this need. It has been heavily inspired by ``Occupy the Cloud'' paper  \cite{PyWren2017}  and
the statement made by a professor of computer graphics at UC Berkeley quoted in that paper: 

\begin{displayquote}
\emph{``Why is there no cloud button?'' He outlined how his students simply wish they could easily ``push a button'' and have their code – existing, optimized, single-machine code – running on the cloud.''}
\end{displayquote}

While serverless computing has been argued to be  cloud computing's next step, moving forward to serverless to meet the above promise is still very challenging. One of the main obstacles resides in the 
serverless programmability model.  Porting existing applications into today's serverless platforms is not a trivial matter. Very often, it requires application redesign, new code development and learning new APIs. 
The same  can be told  for building analytics applications over existing severless offerings, which, as we have seen in \S\ref{sec:related-work},  end up in \servermix systems. 

Consequently,  our primary goal is \textit{to create a serverless data analytics platform, which  ``democratizes Big Data”~by overly simplifying the overall life cycle and cloud programming models}
of data analytics.   To this end, the three-year CloudButton research project ($2019$-$2022$)  will be undertaken as a collaboration between key industrial partners such as IBM, RedHat, and Atos,  
and academic partners such as Imperial College London, Institute Mines T\'{e}l\'{e}com/T\'{e}l\'{e}com SudParis, and Universitat Rovira i Virgili.  To demonstrate the impact of the project, we target 
two settings with large data volumes: \textit{bioinformatics} (genomics, metabolomics) and \textit{geospatial data} (LiDAR, satellital), through institutions and companies such as  EMBL, Pirbright Institute, 
Answare and Fundaci\'{o}n Matrix. 


 CloudButton defines the following goals:
\begin{itemize}
\item \textbf{A high-performance serverless compute engine for Big Data:} This is the foundational
technology for the CloudButton platform to overcome the current limitations of existing
serverless platforms. It includes various  extensions to: 
\begin{enumerate*}[label=(\roman*)]
\item  support stateful and highly performant execution of serverless tasks;
\item  optimized elasticity and operational management of functions built on new locality-aware scheduling algorithms;
\item efficient QoS management of containers that host serverless functions; and 
\item a serverless execution framework to support typical dataflow models.
\end{enumerate*}
This goal will mainly focus on the \emph{scheduling} trade-off aspect.

\item  \textbf{Mutable, shared data  support in serverless computing:} To simplify the transitioning from
sequential to (massively-)parallel code, CloudButton will design a novel middleware to allow the  quickly spawning and easy sharing of mutable data structures in severless platforms. 
This middleware will: 
\begin{enumerate*}[label=(\roman*)]
\item offer an easy-to-use programming framework to handle state on FaaS platforms;  
\item provide dynamic data replication and tunable consistency to match the performance requirements of serverless data analytics;  and
\item integrate these new features into in-memory data grids for superior performance.
\end{enumerate*}
This goal will explore the \emph{dissagregation} area of the IDS trade-off.
\item  \textbf{Novel serverless cloud programming abstractions:} CloudButton will provide a new programming 
model for serverless cloud infrastructures that can express a wide range of existing data-intensive applications with minimal changes. The programming
model should at the same time:
\begin{enumerate*}[label=(\roman*)]
\item preserve the benefits of a serverless execution model in terms of resource efficiency, performance, scalability and fault tolerance; and 
\item explicit support for stateful functions in applications, while offering guarantees with respect to the consistency and durability of the state.
\end{enumerate*}
This goal will mostly concentrate on the \emph{isolation} aspect of the IDS trade-off. 
\end{itemize}

In what follows, we will delve deeper into each of these goals, highlighting  in more detail the advance of each one with respect to 
the state of the art. 

\subsection{High-performance serverless runtime}

In many real-life cloud scenarios, enterprise workloads cannot be straightforwardly moved to a centralized public cloud due to the cost, regulation, latency and bandwidth, 
or a combination of these factors. This forces enterprises to adopt a hybrid cloud solution. However, current serverless frameworks are centralized. That is, out-of-the-box, 
they are unable to leverage computational capacity available in multiple locations. Another gap in the current serverless computing implementations is their obliviousness 
to cloud functions' QoS. In fact, serverless functions are treated uniformly, even though performance and other non-functional requirements might differ dramatically 
from one workload to another. 

Big Data analytics pipelines (a.k.a. analytics workflows) need to be efficiently orchestrated. There exists many serverless workflows orchestration tools~\cite{fission-flows, argo, airflow, brigade}, 
ephemeral serverless composition frameworks~\cite{composer}, and stateful composition engines~\cite{StepFunctions,durablefunc}. To the best of our knowledge, workflow orchestration tools 
treat  FaaS runtimes as black boxes that are oblivious to the workflow structure. This approach, while gaining in portability, has drawbacks related to performance, because an important information 
related to scheduling of the serverless functions that can be inferred from the workflow structure is not shared with the FaaS scheduler.  

A major issue with FaaS, which is exacerbated in a multi-stage workflow, is the data shipment architecture of FaaS. Usually, the data is located in a separate storage service, such as 
Amazon S3 or IBM COS, and shipped for computation to the FaaS cluster. Likewise, the output of the previous FaaS function(s) that might serve as input to the subsequent function(s) in the flow is re-shipped anew and, in 
general, FaaS functions are not scheduled with data locality in mind, even though data locality can be inferred from the workflow structure.

Further, and to the best of our knowledge,  none of the existing workflow orchestration tools is serverless in itself. That is, the orchestrator is usually a stateful, always-on service. 
This is not necessarily the most cost-efficient approach for long running big data analytics pipelines, which might have periods of very high peakedness requiring massive parallelism 
interleaved with long periods of inactivity. 

Last but not least, in a \servermix model, which realistically assumes both serverless and non-FaaS components, the cost effectiveness of the whole analytics pipeline depends on the time utilization 
of each component. In this sense, ``smart'' provisioning algorithms that help to select the right proportion of FaaS vs. non-FaaS components will be of great value to improve the overall cost-efficiency
(see, for instance, \cite{locus}).

In CloudButton, we will address the above challenges as follows:
\begin{itemize}
    \item \textbf{Federated FaaS model}: CloudButton will exploit k$8$s federation architecture to provide a structured multi-clustered FaaS run time to facilitate analytics pipelines spanning 
		multiple k$8$s clusters. The FaaS frameworks that we plan to extend to fit the federated architecture are CNCF Knative and Apache OpenWhisk with public cloud FaaS offerings pluggable 
		to the system, albeit with much less control over scheduling, as explained above.
    \item \textbf{SLA, QoS and scheduling}: programmers will be enabled to specify desired QoS levels for their functions. These QoS constraints will be enforced by a specialized scheduler (implemented via the k$8$s 
		custom scheduler framework). This scheduler will also take the structure of a workflow into account and use this information to improve performance by e.g., pre-warming containers, pre-fetching data, 
		caching data from previous stages, and migrating to remote clusters when local capacity is exhausted. SLAs corresponding to specific QoS levels will be monitored and enforced.
    \item \textbf{\servermix workflow orchestration}: we will construct a serverless orchestration framework for \servermix analytics pipelines by extending mature native k$8$s tools, e.g., Argo~\cite{argo}. Tasks 
		in the \servermix workflow might include massively parallel serverless computations carried out in PyWren~\cite{ibmpywren}. The orchestrator will take care of PyWren invocation restarts, traffic shaping 
		(e.g., how many invocations per time unit), completion handling, etc., moving the burden of orchestration from the PyWren client to the platform and leaving PyWren with the application related tasks, such as smart data partitioning.
    \item \textbf{Operational efficiency}: an operations cost-efficiency advisor will track the time utilization of  each \servermix component and submit recommendations on its appropriate usage. For
		example, a component, which is in constant use, might be more cost-efficiently provided and operated as a serverful one rather than FaaS, while a component utilized below some break-even point depending on the cost of the 
		private infrastructure and/or public cloud services can be more efficiently operated using the serverless approach. 
\end{itemize}

\subsection{Mutable shared data for serverless computing}

In the context of Big Data applications, workloads abide by what we would call \emph{storm computing}, where thousands of serverless functions happen in a brief period of time.
From a storage perspective, this requires the ability to scale abruptly the system in order to be on par with demand.
To achieve this, it is necessary to decrease startup times (e.g., with unikernels~\cite{Manco:2017}) and consider new directions for data distribution 
(e.g., Pocket \cite{pocket}, which uses a central directory and provisions storage nodes in advance).

Current serverless computing platforms outsource state management to a dedicated storage tier (e.g., Amazon S$3$).
This tier is agnostic of how data is mutated by functions, requiring data (de-)serialization in serverless functions.
Such an approach is cumbersome for complex data types, decreases code modularity and re-usability, and increases the cost of 
manipulating large objects. In contrast, we advocate that the storage tier should support in-place modifications \textemdash similarly to what DBMS systems 
offer with stored procedures ~\cite{stonebraker:1991}.

Serverless computing infrastructures have additional key requirements on the storage tier to permit efficient Big Data manipulations.
These are:
\begin{itemize}
\item \textit{Fast access (sub-millisecond) to ephemeral mutable data:} to support iterative and  stateful computations (e.g., ML algorithms);
\item \textit{Fine-grained operations to coordinate concurrent function invocations} (similarly to coordination kernels such as Apache Zookeeper \cite{zookeeper}); and
\item \textit{Dependability:} to transparently support failures in both storage and compute tiers.
\end{itemize}

In CloudButton, we envision to tackle the above challenges by designing a novel storage layer for stateful serverless computation.
Our ultimate goal is to simplify to the minimal expression the transitioning from sequential to massively-parallel code.
This requires to advance the state of the art on several vital questions in data storage and distributed algorithms.
Below, we list the features that we aim to achieve in the storage system.
\begin{itemize}
\item \textbf{Language support for mutable shared data.}
  The programmer can declare mutable shared data types in a piece of serverless code.
  This declaration is integrated transparently to the programming language (e.g., with the help of annotations).
  The storage tier knows the data types, allowing in-place mutations.
  Furthermore, these data types are composable and sharded transparently for performance.
\item \textbf{Tunable data consistency.}
  Shared data objects are distributed and replicated across the storage layer.
  Strong consistency maintains application's sequential invariants but performance generally suggests to use weaker consistency models \cite{WadaFZLL11,Attiya:1991}.
  To reconcile ease of programming and performance, developers can \emph{degrade} data consistency.
  This degradation is controlled at the level of individual object and integrated to the language support.
\item \textbf{Just-right synchronization.}
  Each object is implemented using state machine replication atop a consensus layer \cite{paxos,raft}.
  This layer is adaptable and self-adjusts to the consistency of each shared data item.
  Doing this, data replicas synchronize only when necessary, transforming consistency degradation into performance.
\item \textbf{In-memory data storage.}
  Shared data is stored in-memory and overflows to external storage (e.g., filesystem, database, etc.) when it is tagged as persistent (e.g., the centroids at the end of a $k$-means clustering job).
  To cope with the short-lived, highly-demanding nature of the workload,
  \begin{itemize}
  \item Data distribution is computed before computation occurs;
  \item Persistent and in-memory data nodes collaborate; and
  \item Sorage adapts replication and locality on-the-fly with the help of an external orchestrator.
  \end{itemize}
\end{itemize}

\subsection{Novel serverless cloud programming abstractions: the CloudButton toolkit}

Containers are the foundation of serverless runtimes, but the abstractions and isolation they offer can be restrictive for many applications. A hard barrier 
between the memory of co-located functions means all data sharing must be done via external storage, precluding data-intensive workloads and introducing 
an awkward programming model. Instantiating a completely isolated runtime environment for each function is not only inefficient, but at odds with how most 
language runtimes were designed. 

This isolation boundary and runtime environment have motivated much prior work. A common theme is optimizing and modifying containers to better suit the task, 
exemplified by SOCK\cite{oakes2018sock}, which makes low level changes to improve start-up times and efficiency. Others have partly sacrificed isolation to achieve better 
performance, for example by co-locating a tenants' functions in the same container \cite{sand}. Also, a few frameworks for building serverless applications have 
emerged \cite{PyWren2017,flint,ttt}. But these systems still require a lot of engineering effort to port existing applications.

Software fault isolation (SFI) has been proposed as an alternative isolation approach, offering memory-safety at low cost \cite{Boucher2018}. Introducing an intermediate representation (IR) to unify the spectrum 
of languages used in serverless has also been advocated \cite{hellerstein2018serverless}. WebAssembly is perfectly suited on both counts. It is an IR built on the principles of SFI, designed for executing 
multi-tenant code \cite{wasm2017}. This is evidenced by its use in proprietary serverless technologies such as CloudFlare Workers \cite{cloudflare} and Fastly's Terrarium \cite{FastlyTerrarium}.

With the CloudButton toolkit, we will build on these ideas and re-examine the serverless programming and execution environment. We will investigate 
new approaches to isolation and abstraction, focusing on the following areas:

\begin{itemize}
    \item \textbf{Lightweight serverless isolation.} By combining SFI, WebAssembly and existing OS tooling we will build a new isolation mechanism, delivering strong security guarantees at a fraction of the cost of containers. This will be the foundation on which we construct the rest of the toolkit.  
    \item \textbf{Efficient localized state.} This new isolation approach allows sharing regions of memory between co-located functions, enabling low-latency parallel processing and new opportunities for inter-function communication. We will build on this to tackle data-intensive workloads and investigate how our scheduling can benefit from co-location.
    \item \textbf{Stateful programming abstractions.} To make CloudButton programming seamless, we will create a new set of abstractions, allowing users to combine stateful middleware with efficient localized state to easily build high-performance parallel applications.
    \item \textbf{Serializable execution state.} WebAssembly's simple memory model makes it easy to serialize and resume a function's execution state. We will create a checkpoint, migrate and restore mechanism to make horizontal scaling across hosts transparent to the user.
    \item \textbf{Polyglot libraries and tooling.} By using a shared IR we can reuse abstractions across multiple languages. In this manner we will build a suite of generic tools to ease porting existing applications in multiple languages, including the CloudButton genomics and geospatial use-cases.
\end{itemize}

\section{Conclusions and future directions}

In this article, we have first analyzed three important architectural trade-offs of serverless computing: disaggregation, isolation, and simple scheduling. We have explained that by relaxing 
those trade-offs, it is possible to achieve higher performance, but also how that loosening can  impoverish important serverless traits  such as elasticity, multi-tenancy 
support, and high resource utilization. Moving the trade-offs to the extremes, we have distinguished between serverful and serverless computing, and we have also introduced the new concept 
of \servermix computing. 

\servermix systems combine serverless and serverful components to accomplish an analytics task. An ideal \servermix system should keep resource provisioning transparent to the user and consider
the cost-performance ratio as first~citizen. Indeed, we have found that most of the related work can be categorized as \servermix, thereby confirming that a deep understanding of  \servermix
designs and their limitations is central for what is being called ``serverless computing'' today. In this context,   we have presented the CloudButton project and posed several open 
research challenges for the future, such as smart scheduling, fine-grained  state  disaggregation, and lightweight  and  polyglot  isolation.

Although we predict that \servermix systems  will  decline  over  time, e.g., because of the irruption of novel serverless services covering the entire cycle of Cloud applications, 
some deployments might persist in certain domains due to powerful reasons such as performance and efficiency, to name a few. For instance, shared memory or computation close to the data will still be required in many HPC programming models like OpenMP. This is just a simple example to illustrate that while $100\%$ serverless analytics systems cannot be built now, advances in the \servermix model will be proven
very valuable.  

We conclude this article with the following research directions to explore to provide truly serverless analytics systems:
\begin{itemize}
\item We believe that providing  APIs for service-level objectives (SLOs) in FaaS platforms will be of great help to attract a critical mass of
workloads. This will raise performance predictability and the ability to provide cost-effective serverless solutions for analytics tasks.  Ideally,
the cloud provider should infer from the application, or user specifications (e.g., via high-level DSLs), the most cost-effective solution that fulfills user 
specified SLOs, combining both FaaS and BaaS services.
\item Another research direction will be to develop a serverless, fast storage service with a rich interface to support a wide variety of application
requirements like: in-place data reductions, fine-grained updates and coordination, and tunable consistency, among others.  This new storage
service should be compliant with the disaggregation principle, and come in ephemeral and durable variants~\cite{berkeleyserverless}.
\item  In addition to lightweight isolation to minimize multi-tenancy overheads, the use of a common intermediate representation (IR) to serve as unified substrate for many languages
will contribute to significantly simplify the programming of serverless computing.  Such IR should be designed for multi-tenant code \cite{wasm2017} 
and support a large variety of languages and FaaS platforms.  WebAssembly is a promising candidate for this endeavor. 
\end{itemize}

\section*{Acknowledgement}
This work has been partially supported by the EU project H2020 ``CloudButton: Serverless Data Analytics Platform'' (825184)
and by the Spanish government TIN2016-77836-C2-1-R. Thanks also due to the Serra Hunter programme from the Catalan Government.

\bibliographystyle{unsrt}  
\bibliography{servermix}

\end{document}